\def\m{\rm \,m}

\def\K{\rm \,K}

\def\W{\rm \,W}

\def\sec{\rm \, sec}
\def\km{\rm \, km}

\documentstyle[11pt,twoside,colloqOHP,epsfig]{article}
\begin{document}

\title{Atmospheric dynamics of Pegasi planets}
\author{Adam P. Showman and Curtis S. Cooper}
\affil{Department of Planetary Sciences, Lunar and Planetary Laboratory,
University of Arizona, Tucson, AZ 85721 [showman@lpl.arizona.edu, curtis@lpl.arizona.edu]}

\begin{abstract}
We present three-dimensional numerical simulations of the atmospheric dynamics
of close-orbiting planets such as HD209458b.  Our simulations show that 
winds of several $\km\sec^{-1}$ and day-night temperature differences reaching
500--$1000\K$ are possible at and above the photosphere.
The circulation takes the form of a broad superrotating (eastward) equatorial jet.
At $\sim$0.1--$1\,$bar, the superrotation blows the hottest regions of the
atmosphere downwind by $\sim60^{\circ}$ of longitude, but at lower pressures the 
temperature pattern tracks the stellar illumination.  We predict factors of several
variation in the infrared flux received at Earth throughout an orbital cycle;
if the photosphere is deep enough ($\ge$50--100 mbar pressure),
the peak infrared emission should lead the time of secondary eclipse by 10 hours
or more.  Dynamics
plays a key role in shaping the spectrum, clouds, chemistry,
and long-term planetary evolution.

\end{abstract}

The past few years have witnessed many observations constraining the physical
properties of extrasolar giant planets with orbital radii less than 0.1 AU
(the ``Pegasi planets'' or hot Jupiters).  Eight such planets have been discovered
to undergo transits.   Two such planets, HD209458b and TrES-1, have also been detected
in thermal emission during the secondary eclipse, and several useful upper limits on
composition, albedo, and thermal emission at various wavelengths have been achieved.
This trend of detections is likely to continue.

A knowledge of atmospheric dynamics will be crucial for understanding these
new observations.  First, the interaction of dynamics with radiative transfer controls
the temperature structure, which shapes the infrared spectrum and lightcurve.
Most current radiative-transfer models adopt radiative-equilibrium conditions
and make arbitrary assumptions about whether the absorbed stellar flux gets redistributed
across the planet or heats only the dayside; however, dynamics can push the
atmosphere far from radiative equilibrium, and the extent of heat redistribution
must be calculated explicitly (and may depend strongly on height).
Second, whether clouds exist depends on the temperatures and locations of
ascent/descent, which is again controlled by the circulation. 
Cloudiness in turn determines the albedo, visible lightcurves, and
--- if high-altitude clouds form --- causes masking of spectral lines. 
Third, circulation may lead to disequilibrium
between CO and CH$_4$, remove condensable species (Na$_2$S, CaTiO$_3$) via nightside
cold trapping, and cause other chemical effects.  Fourth, the atmospheric heat
engine produces enormous kinetic energy, which, if transported deep enough,
may affect the interior evolution.  This has been suggested as a possible mechanism
for producing the large radius of HD209458b, for example (Guillot and Showman 2002,
Showman and Guillot 2002).  

The intense starlight incident upon the surface of Pegasi planets leads to
a deep radiative zone extending from the top of the atmosphere to pressures of
$\sim1000\,$bars (e.g., Guillot et al. 1996, Guillot and Showman 2002, Burrows et al. 2003,
Chabrier et al. 2004), and any observable weather occurs in this radiative zone. 
The fast spindown times for Pegasi planets implies that these planets should be
in near-synchronous rotation (3.5 days for HD209458b)(Guillot et al. 1996, 
Showman and Guillot 2002).  This rotation rate implies that rotation is important 
but not dominating: for  $\km\sec^{-1}$ winds, the Rossby number is $\sim1$.  
These estimates imply dynamical length scales 
(the Rossby deformation radius and Rhines length) of order a 
planetary radius.  As a result, any jets and gyres that exist 
should be global in scale.  This contrasts with the case of Jupiter,
where these length scales are only $2$--10\% of the planetary radius
and --- as a result --- the dominant jets and vortices are much 
smaller than a planetary radius.  Pegasi planets should therefore 
have physical appearances that differ greatly from Jupiter and 
Saturn.

\begin{figure}
\includegraphics[scale=1.0]{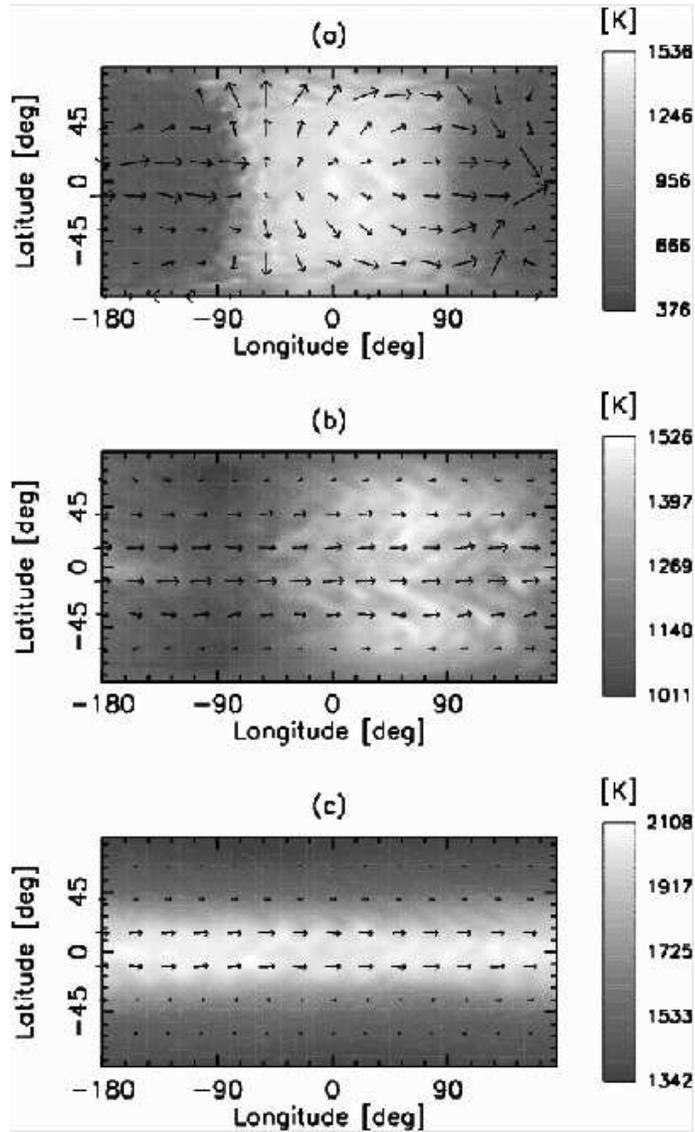}
\caption{Temperature (greyscale) and winds (arrows) at pressures
of 2 mbar, 220 mbar, and 20 bars at 5000 Earth days of simulated time.
Peak winds are 9.2, 4.1, and $2.8\km\sec^{-1}$ from top to bottom,
respectively. Heating occurs on the dayside (longitudes $-90^{\circ}$
to $90^{\circ}$) and cooling occurs on the nightside (longitudes
$-180^{\circ}$ to $-90^{\circ}$ and $90^{\circ}$ to $180^{\circ}$).
The substellar point is at $0^{\circ}$ latitude, $0^{\circ}$ longitude.}
\end{figure}

Here we describe our recent work on the atmospheric
circulation of Pegasi planets; the presentation describes and extends Cooper and
Showman (2005), to which we refer the reader for details.
We performed global, three-dimensional 
numerical fluid simulations using the ARIES/GEOS dynamical 
core (Suarez and Takacs 1995) in a domain extending from $1\,$mbar---$3\,$kbar.
The simulations adopted the primitive equations, which are a simplified form 
of the Navier-Stokes equations
valid for statically stable atmospheres that are vertically thin compared to
their horizontal extent.  We used parameters for HD209458b and assumed that the planetary 
interior is in synchronous rotation with the orbital period.  The nominal resolution
is $72\times45$ in longitude and latitude with 40 vertical levels.  In the simulations,
the dynamics are driven solely by the imposed day-night heating contrast.  Rather than solving
the radiative transfer explicity, we adopted a thermodynamic heating 
rate (in $\K\sec^{-1}$) of $(T_{\rm eq} - T)/\tau_{\rm rad}$, where $T_{\rm eq}$
is the specified radiative-equilibrium temperature profile (hot on the dayside 
and cold on the nightside), 
$T$ is the actual temperature, and $\tau_{\rm rad}$
is the radiative-equilibrium timescale (a function of pressure).  The vertical
structure of $T_{\rm eq}$ and $\tau_{\rm rad}$ were taken from Iro et al. (2005); 
the day-night difference in $T_{\rm eq}$ was a free parameter that we varied
from from 100---1000 K.

Figure~1 shows the temperature (greyscale) and winds (arrows) for three
layers (2 mbar, 200 mbar, and 20 bars from top to bottom, respectively)
after a simulated time of 5000 Earth days.  The imposed heating contrast
leads to winds exceeding several $\km\sec^{-1}$.  By 5000 days the simulation
has approximately reached a statistical steady state at pressures less
than 3 bars, although the winds continue to increase at deeper levels.
At the top (2 mbar), the radiative time constant is $\sim1\,$hour (much less
than the advection time), so the
hot regions remain confined to the dayside.  The temperatures are in
near-radiative-equilibrium, with day-night temperature differences of $\sim1000\K$.
At 200 mbar, close
to the expected photosphere if the planet lacks high-altitude clouds,
a broad $\sim4\km\sec^{-1}$ eastward superrotation develops.  Here,
the radiative time constant, $\sim10^5\sec$, is comparable to the time
needed to advect air across a planetary radius.
The circulation therefore blows the hottest regions of
the atmosphere downwind from the substellar point by $\sim60^{\circ}$
degrees of longitude.  Temperature differences reach $\sim500\K$ at this level.

\begin{figure}
\includegraphics[scale=0.6]{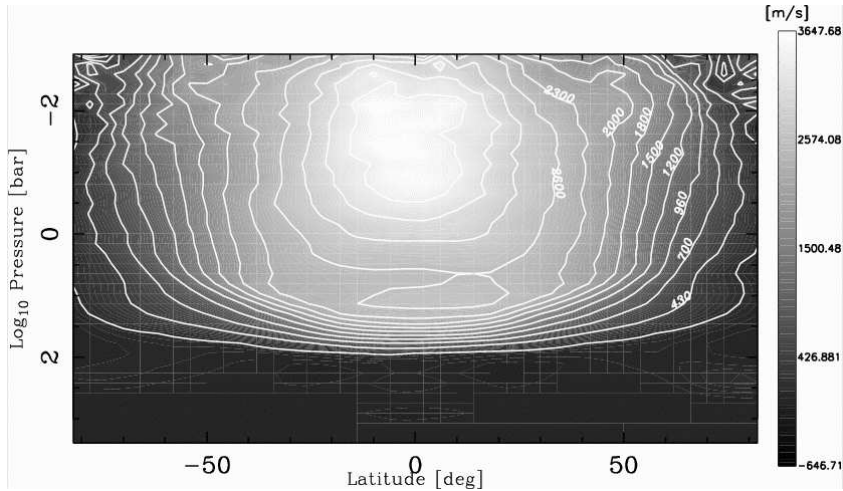}
\caption{Longitudinally averaged east-west winds versus latitude and pressure from
the simulation in Fig.~1. Positive values are eastward.}
\end{figure}

The development of a broad, superrotating (eastward) jet, with large day-night
temperature differences at $p<1\,$bar, is a robust feature in all our simulations.  
We performed a variety of simulations using radiative-equilibrium temperature
profiles from Burrows et al. (2003), Chabrier et al. (2004), or Iro et al. (2005);
and using radiative-equilibrium day-night temperature differences of 1000,
750, 500, 250, or $100\K$.  We even performed simulations whose initial
condition contained a broad {\it westward} equatorial jet extending from $\sim2\,$bars to
the top of the domain.  All of these simulations developed strong eastward jets 
resembling that in Fig.~1.  Furthermore, these results agree with Showman and Guillot (2002), 
who also obtained broad eastward jets in every one of 
their simulations using a different numerical code.
This gives us confidence that eastward flow is a robust result, 
at least within the context of our adopted input parameters.   What all these
simulations have in common are short radiative time constants at pressures
$\le1\,$bar, which allow the development of longitudinal temperature
variations that are essentially a large-amplitude thermal tide.
We speculate that, as has been suggested for Venus,
this tide induces the superrotation by pumping eddy energy and 
eastward momentum upward and equatorward.  (In absence of such eddy effects,
the equatorial flow would be westward.)  

\begin{figure}
\includegraphics[scale=0.6]{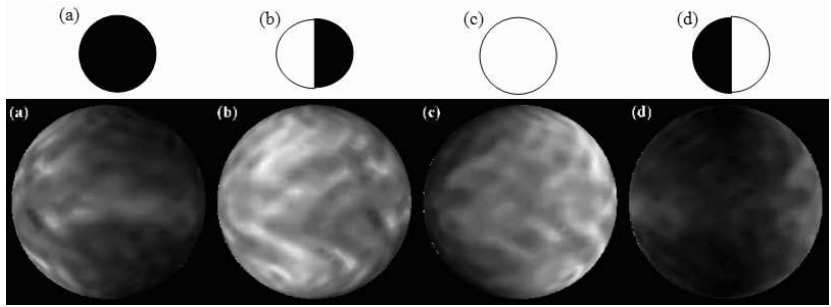}
\caption{Infrared brightness of HD209458b as viewed at Earth during (a) transit,
(b) one-quarter period after tranit, (c) secondary eclipse, and (d) 
one-quarter period after secondary eclipse.  The planetary rotation axes are 
vertical, with the superrotating jet seen in Fig.~1 going from left to right
in each panel.  The smaller schematic
globes in the top row illustrate the illumination of the planet by sunlight, 
as viewed at Earth, during these same phases. Time increases from left to
right.}
\end{figure}

Our results differ from the
one-layer shallow-water calculations of Cho et al. (2003), which produce
westward equatorial flow.  However, shallow-water turbulence invariably
produces westward flow even for planets such as Jupiter and Saturn whose
equatorial jets are eastward (e.g., Cho and Polvani 1996, Iacono et al. 1999;
see Vasavada and Showman 2005 for a review).  This
feature seems to result from the exclusion of three-dimensional processes
in the shallow-water equations.

The patterns in Fig.~1 have implications for the infrared lightcurve
of the planet throughout its orbit, as shown in Figs.~3 and 4. The
globes in Fig.~3 (bottom row) show the infrared brightness at four phases
assuming the planet emits as a blackbody from the 220 mbar level.
The globes in the top row show the illumination as viewed from
Earth during these same phases.  The key point is that, in the
absence of winds, the temperature pattern would follow the illumination
(i.e., the infrared appearance would also correspond to the top
row of globes).  The differences between the idealized illumination
patterns and the simulated brightness patterns result solely from
atmospheric dynamics.  

Figure~4 (left) shows the corresponding
lightcurve assuming the photosphere is at the 220 mbar level, as might
be expected for a cloud-free planet.  Because the hot regions become offset 
from the substellar point at this pressure, the model predicts that the 
planet will radiate its maximum
infared flux toward Earth $\sim14\,$hours {\it before} the secondary
eclipse (rather than immediately around the time of secondary eclipse
as would occur without winds).  This effect could allow an observational
determination of wind direction --- if the winds are eastward, the peak
fluxes would lead the eclipse, whereas if the winds are westward, peak
fluxes would lag the eclipse.

\begin{figure}
\includegraphics[scale=0.55]{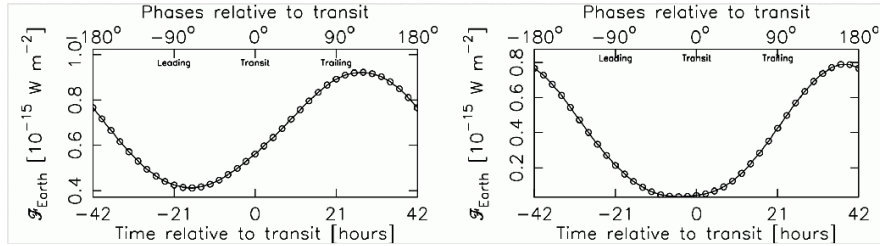}
\caption{Predicted infrared lightcurve for HD209458b assuming blackbody emission
from the 220 mbar level (left) or the 10 mbar level (right).}
\end{figure}

Nevertheless, uncertainties exist regarding the emission level.
The lower-than-expected Na (Charbonneau et al. 2002) and null detection
of CO (Deming et al. 2005, Richardson et al. 2003) on HD209458b
could result from a high-altitude cloud at a few mbar pressure.  If
the cloud optical depth exceeds unity, then the primary infrared emission to space
occurs from the cloud altitude rather than from the deeper levels expected 
for a cloud-free planet.   Figure~4 (right) shows
the lightcurve for the case of emission from the 10-mbar level.  
Because the radiative time constant
is short at these pressures (Iro et al. 2005), the offset shown
in Fig.~4 (left) has largely disappeared.  The magnitude of the flux
differences has increased from $\sim$two-fold (Fig.~4, left) to
eightfold (Fig.~4, right).  Emission from such high altitudes would
largely mask the signature of winds.  It is possible that some
planets have high-altitude clouds while others do not (HD209458b
and TrES-1 may represent these cases; Fortney et al. 2005), so
a range of infrared-lightcurve behaviors is to be expected among
real planets.

\begin{figure}
\includegraphics[scale=0.42]{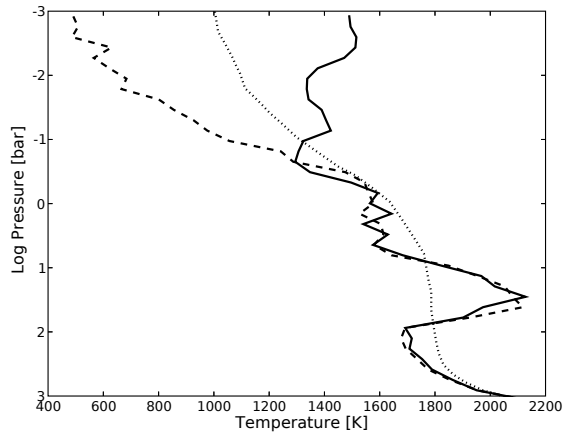}
\caption{Solid and dashed curves show the temperature profile at the
substellar and antistellar points, respectively, for the simulation in
Fig.~1.  The dotted line shows the globally averaged 
radiative-equilibrium profile from Iro et al. (2005).  Note the formation
of a dayside inversion layer at pressures less than 0.3 bars even though none 
exists in radiative equilibrium.}
\end{figure}

Dynamics can push the atmosphere far from radiative equilibrium,
and this may have implications in explaining the existing Spitzer
IRAC data at 4.5 and $8\,\mu$m  for TrES-1 (Charbonneau et al. 2005).  
Current radiative-equilibrium models cannot easily explain the data: 
if they explain the $4.5\,\mu$m flux, then they do not predict enough flux at 
$8\,\mu$m (Fortney et al. 2005, Burrows et al. 2005, 
Seager et al. 2005, Barman et al. 2005).  Part of the problem is that
8-$\mu$m photons are emitted from higher altitude, where radiative-equilibrium
models predict colder temperatures.  One solution is to invoke
a temperature inversion so that the emission region for 8-$\mu$m
photons is hotter than for 4.5-$\mu$m photons.  Fortney et al. (2005) 
accomplished this by adding an {\it ad hoc} heat source, which lead
to an improved fit to the Spitzer data.   A key point is that dynamics 
can {\it naturally} produce such a dayside temperature inversion, even when
no such inversion would exist in radiative equilibrium (Fig.~5).  The
inversion occurs because of the upward-decreasing radiative time
constant: as air columns superrotate from nightside to dayside, the air
at the top warms much more rapidly than air at the bottom, producing
an inversion.  No {\it ad hoc} heat sources need be invoked.

\begin{figure}
\includegraphics[scale=0.42]{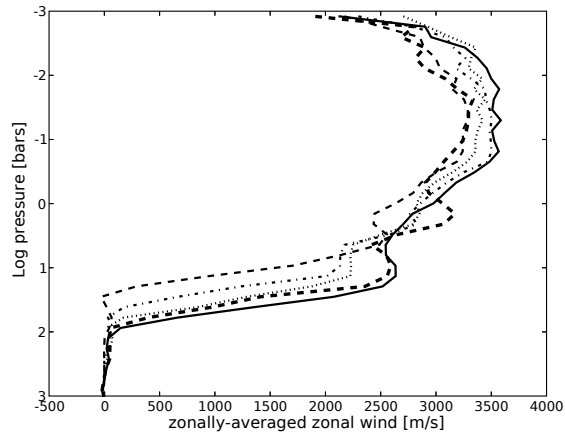}
\caption{Time evolution of the longitudinally averaged equatorial east-west 
winds from the simulation in Fig.~1.  Thin dashed, dash-dot, dotted,
thick dashed, and solid curves show profiles at 1000, 2000, 3000, 4000, and 5000
Earth days of simulated time.  Note the gradual downward penetration of
winds from 10--100 bars over the course of the simulation.}
\end{figure}

The large radius of HD209458b has remained a puzzle, particularly
given that the other 7 known transiting planets have radii in
agreement with evolution calculations (e.g., Guillot 2005).  Showman
and Guillot (2002) and Guillot and Showman (2002) suggested that
mechanical energy produced by the atmospheric heat engine could
be advected into the interior, where it could be dissipated and might
provide a source of heat that would slow the contraction.  In the
current simulations, all of the heating/cooling (which is the sole energy
source in the simulations) occurs at pressures $\le10\,$bars.  Nevertheless,
the simulations gradually develop strong winds at pressures $>10\,$bars,
which implies that kinetic energy is transported downward from the heated
regions into the interior.  Figure~6 shows the evolution
of longitudinally averaged winds at the equator over time.  The
winds at pressures $\le1\,$bar rapidly reach a quasi-steady equilibrium,
but the winds from 10--100 bars increase throughout the simulation
(this implies a large increase in total kinetic energy because that
layer contains ten times more mass than the entire overlying atmosphere).
The build-up of winds in Fig.~5 corresponds to a downward kinetic
energy flux of $\sim10^3\W\m^{-2}$, which is 10--100 times greater
than the intrinsic flux predicted in evolution models (e.g., Guillot
and Showman 2002, Burrows et al. 2003, Chabrier et al. 2004).  More work is needed
to determine the fate of this energy, but it suggests that atmospheric
circulation could affect the long-term evolution.

\acknowledgments{
This work was supported by NSF grant AST-0307664 and NASA GSRP NGT5-50462.
}

\end{document}